# A Novel Mobility Model to Support the Routing of Mobile Energy Resources

Wei Wang, *Member, IEEE*, Xiaofu Xiong, *Member, IEEE*, Chao Xiao, and Bihui Wei

*Abstract*—Mobile energy resources (MERs) have received increasing attention due to their effectiveness in boosting the power system resilience in a flexible way. In this paper, a novel mobility model for MERs is proposed, which can support the routing of MERs to provide various services for the power system. Two key points, the state transitions and travel time of MERs, are formulated by linear constraints. The feasibility of the proposed model, especially its advantages in model size and computational efficiency for the routing of MERs among many nodes with a small time span, is demonstrated by a series of tests.

*Index Terms*—Mobile energy resources, power system resilience, mobility model, routing, linear constraint.

## I. Introduction

MOBILE energy resources (MERs) can act as "first aid boxes" to rapidly restore and maintain electric service to customers when the power system suffers blackouts resulting from, *e.g.*, severe natural events and cyber attacks [1], [2], [3].

To date, rather little research has studied the routing of MERs, whereas its effectiveness for boosting the resilience and economics of the power system has been highlighted [4]-[9]. The routing of MERs for the above purposes is always formulated as a programming problem, and the model depicting the time- and space-related travel behavior of MERs is exactly regarded as the kernel component of that programming.

From a review of the relevant research, two major mobility models have been used to formulate the travel behavior of MERs to support their routing. 1) The time-space network (TSN) [5], [6] uses the arcs between nodes to represent all possible behaviors of MERs in each time span. 2) In addition, two simply formulated mobility models with almost the same structure were given in [7] and [8], respectively, as part of the constraints in programming for routing MERs, and herein let us call them sliding window-based model (SWBM). The SWBM depicts that the parking label of an MER (*i.e.*, the parking state of 1) cannot transit from one node to another unless the time interval exceeds the travel time between the two nodes.

However, there are some inherent drawbacks for both models. For either TSN or SWBM, the model size increases greatly with routing scope, reflected as the dramatic increase in the number of binary variables or constraints with the square of the number of nodes that support the connection of MERs, even though some improvement has been made for TSN in [9]. Moreover, a small time span will also make the two models much more large, whereas a smaller span is better for the routing quality.

Therefore, TSN and SWBM may experience degraded efficiency when they are used for routing of MERs among many nodes, while the routing is truly an urgent task for, *e.g.*, electric service restoration issue where MERs are generally scheduled to restore the loads as many and as soon as possible in the whole power system [5], [7], [8], [9]. Time for solving the model delays the execution of the obtained optimal schedule of tasks for the restoration, and thus the utility always desires the model can be solved as quickly as possible. We might adopt a rough time span to improve the efficiency at the large sacrifice of routing quality; however, making such a trade-off is not easy.

To bridge this gap, this paper proposes a novel mobility model to support the routing of MERs. The model formulates the travel behavior of MERs by linear constraints and thus can be well embedded as a kernel part in a mixed-integer linear programming (MILP) or others for the issues involving routing of MERs that can be accurately solved by off-the-shelf solvers.

## II. A Mobility Model for MERs

### A. Constraints of State Transition

MERs can travel among the nodes (or buses) of power grid and exchange power with it, with the support of necessary facilities, *e.g.*, charging stations. We use sets $\mathcal{N}$, $\mathcal{M}$, $\mathcal{T}$ to represent the nodes of power system that support the connection of MERs, the MERs, and the time spans for scheduling, respectively; and $\mathcal{T}=\{0, 1, 2, \ldots, D\}$. Two binary variables are defined to denote the parking and traveling states of MER: $x_{j,i,t}$, which is equal to 1 if MER $j$ is parked at node $i$ during time span $t$ and 0 otherwise; and $v_{j,i,t}$, which is equal to 1 if MER $j$ is traveling to node $i$ during time span $t$ and 0 otherwise. Clearly, there is only one state anytime for an MER, as formulated by

$$\sum_{i \in \mathcal{N}} x_{j,i,t} + \sum_{i \in \mathcal{N}} v_{j,i,t} = 1, \forall j \in \mathcal{M}, t \in \mathcal{T} \quad (1)$$

Without loss of generality, we can give the representative segments of the parking and traveling state sequences that we expect in Table I. Assume that, over the period $[t_1+2, t_1+4]$, MER $j$ is traveling from node $i_1$ to $i_2$, so $v_{j,i_2,t}$ owns the traveling label, shown by the blue '1'. As a result, the parking label, shown by the red '1', transits from $x_{j,i_1,t}$ to $x_{j,i_2,t}$.

Based on Table I, we can clearly infer that the parking state of an MER does not change unless a new travel starts or ends, and an idea is then conceived: perhaps we can determine its parking state transitions (PST) from its traveling state transitions (TST). Then, let us constrain the PST by (2) that includes $D_{j,i,t}$ and $U_{j,i,t}$ to be determined, and define '$\Delta_{(1)j,i,t}$' and '$\Delta_{(2)j,t}$' in (3) to denote the TST. According to that idea, we expect to find a way to formulate $D_{j,i,t}$ and $U_{j,i,t}$ by $\Delta_{(1)j,i,t}$ and $\Delta_{(2)j,t}$ in the linearized form to obtain the definite form of (2), as expressed by (4), where $a_1$, $b_1$, $c_1$, $a_2$, $b_2$, and $c_2$ are coefficients to be determined by the following content.

$$x_{j,i,t} - D_{j,i,t} \leq x_{j,i,t+1} \leq x_{j,i,t} + U_{j,i,t}, \forall i \in \mathcal{N}, j \in \mathcal{M}, t \in \mathcal{T} \setminus \{D\} \quad (2)$$

TABLE I
REPRESENTATIVE SEGMENTS OF THE EXPECTED STATES FOR MER $j$

| Time $t$ | Parking state $x_{j,1,t} \ldots x_{j,i_1,t} \ldots x_{j,i_2,t} \ldots x_{j,N,t}$ | | | | Traveling state $v_{j,1,t} \ldots v_{j,i_1,t} \ldots v_{j,i_2,t} \ldots v_{j,N,t}$ | | | | $S_{j,t}$ | $R_{j,t}$ |
|---|---|---|---|---|---|---|---|---|---|---|
| ⋮ | ⋮ | ⋮ | ⋮ | ⋮ | ⋮ | ⋮ | ⋮ | ⋮ | ⋮ | ⋮ |
| $t_1$ | 0 | 1 | 0 | 0 | 0 | 0 | 0 | 0 | 0 | 0 |
| $t_1+1$ | 0 | 1 | 0 | 0 | 0 | 0 | 0 | 0 | 0 | 0 |
| $t_1+2$ | 0 | 0 | 0 | 0 | 0 | 0 | 1 | 0 | $T_{j,i_1 i_2}(3)$ | $T_{j,i_1 i_2}(3)$ |
| $t_1+3$ | 0 | 0 | 0 | 0 | 0 | 0 | 1 | 0 | 0 | $T_{j,i_1 i_2}-1(2)$ |
| $t_1+4$ | 0 | 0 | 0 | 0 | 0 | 0 | 1 | 0 | 0 | $T_{j,i_1 i_2}-2(1)$ |
| $t_1+5$ | 0 | 0 | 1 | 0 | 0 | 0 | 0 | 0 | 0 | 0 |
| $t_1+6$ | 0 | 0 | 1 | 0 | 0 | 0 | 0 | 0 | 0 | 0 |
| ⋮ | ⋮ | ⋮ | ⋮ | ⋮ | ⋮ | ⋮ | ⋮ | ⋮ | ⋮ | ⋮ |

TABLE II
THE FEASIBLE STATE TRANSITIONS AND THE EXPECTED $D_{j,i,t}$ AND $U_{j,i,t}$

| $t$ | PST $(x_{j,i,t} \to x_{j,i,t+1})$ | Required $D_{j,i,t}$ | Required $U_{j,i,t}$ | $D_{j,i,t}^*$ | $U_{j,i,t}^*$ | TST $\Delta_{(1)j,i,t}$ | TST $\Delta_{(2)j,t}$ |
|---|---|---|---|---|---|---|---|
| $t_1+1$ | 1→0 (for $i=i_1$) | ≥1 | ≥−1 | ≥1 | ≥0 | 0 | −1 |
|  | 0→0 (for $i \neq i_1$ or $i_2$) | ≥0 | ≥0 | | | | |
| $t_1+1$ | 0→0 (for $i=i_2$) | ≥0 | ≥0 | ≥0 | ≥0 | −1 | −1 |
| $t_1+4$ | 0→1 (for $i=i_2$) | [−1,0) | ≥1 | [−1,0) | ≥1 | 1 | 1 |
| $t_1+4$ | 0→0 (for $i \neq i_2$) | ≥0 | [0,1) | ≥0 | [0,1) | 0 | 1 |
| $t_1$ | 0→0 (for $i \neq i_1$) | ≥0 | [0,1) | [0,1) | [0,1) | 0 | 0 |
|  | 1→1 (for $i=i_1$) | [0,1) | ≥0 | | | | |
| $t_1+2$, $t_1+3$ | 0→0 (for all $i$) | ≥0 | ≥0 | [0,1) | [0,1) | 0 | 0 |
| $t_1+5$ | 0→0 (for $i \neq i_2$) | ≥0 | [0,1) | [0,1) | [0,1) | 0 | 0 |
|  | 1→1 (for $i=i_2$) | [0,1) | ≥0 | | | | |

$$\Delta_{(1)j,i,t} = v_{j,i,t} - v_{j,i,t+1}, \forall i \in \mathcal{N}, j \in \mathcal{M}, t \in \mathcal{T} \setminus \{D\} \quad (3a)$$

$$\Delta_{(2)j,t} = \sum_{i \in \mathcal{N}} v_{j,i,t} - \sum_{i \in \mathcal{N}} v_{j,i,t+1}, \forall j \in \mathcal{M}, t \in \mathcal{T} \setminus \{D\} \quad (3b)$$

$$\begin{bmatrix} D_{j,i,t} \\ U_{j,i,t} \end{bmatrix} = \begin{bmatrix} a_1 & b_1 \\ a_2 & b_2 \end{bmatrix} \cdot \begin{bmatrix} \Delta_{(1)j,i,t} \\ \Delta_{(2)j,t} \end{bmatrix} + \begin{bmatrix} c_1 \\ c_2 \end{bmatrix} \quad (4)$$

From Table I, we can list all of the feasible state transitions of an MER in Table II, including the PST in the 2nd column and the TST in the 7th and 8th two columns. The intervals of $D_{j,i,t}$ and $U_{j,i,t}$ required to realize each of the PST by (2) are given accordingly in the 3rd and 4th columns. Then, in the 5th (or 6th) column, $D_{j,i,t}^*$ (or $U_{j,i,t}^*$) represents the interval of $D_{j,i,t}$ (or $U_{j,i,t}$) required for all the PST under a pair of $\Delta_{(1)j,i,t}$ and $\Delta_{(2)j,t}$, i.e., under the same TST, which takes the intersection among the required intervals of $D_{j,i,t}$ (or $U_{j,i,t}$) of the included PST. In addition, the effect of (1) is considered when we determine the above intervals of $D_{j,i,t}$ and $U_{j,i,t}$. For example, for the first PST in Table II (i.e., $t=t_1+1$ and $i=i_1$), we should make $x_{j,i,t+1}=0$ while $x_{j,i,t}=1$. However, this is naturally realized only due to (1) while we note that $v_{j,i_2,t+1}=1$; we just need to ensure that (2) is not contradictory to $x_{j,i,t+1}=0$, and thus, we take $D_{j,i,t} \geq 1$ and $U_{j,i,t} \geq -1$.

Then, the determination of the coefficients in (4) can be described as that, for all five pairs of $\Delta_{(1)j,i,t}$ and $\Delta_{(2)j,t}$ in Table II, by (4), we should make $D_{j,i,t}$ (or $U_{j,i,t}$) always within the intersected interval $D_{j,i,t}^*$ (or $U_{j,i,t}^*$) to satisfy all the required intervals in the 3th (or 4th) column, and thus, all the PST can be ensured by (2) and (1). Thus, the feasible coefficients can be determined by the two programming models in (5), in which the constraints are shrunk slightly to simply transform the strict inequalities to non-strict ones for compiling and solving [10].

$$\begin{aligned}
&\min\ f_D(a_1,b_1,c_1) \\
&\text{s.t.: } -b_1 + c_1 \geq 1.2 \\
&\quad -a_1 - b_1 + c_1 \geq 0.2 \\
&\quad -0.8 \leq a_1 + b_1 + c_1 \leq -0.2 \\
&\quad b_1 + c_1 \geq 0.2 \\
&\quad 0.2 \leq c_1 \leq 0.8
\end{aligned} \quad \text{and} \quad \begin{aligned}
&\min\ f_U(a_2,b_2,c_2) \\
&\text{s.t.: } -b_2 + c_2 \geq 0.2 \\
&\quad -a_2 - b_2 + c_2 \geq 0.2 \\
&\quad a_2 + b_2 + c_2 \geq 1.2 \\
&\quad 0.2 \leq b_2 + c_2 \leq 0.8 \\
&\quad 0.2 \leq c_2 \leq 0.8
\end{aligned} \quad (5)$$

In fact, any feasible solution of (5) can be satisfactory. Thus, we do not restrict what the objective functions $f_D$ and $f_U$ are, but only require that they be linear for the simplicity to solve the two models in (5). For example, we can set $f_D=a_1+b_1+c_1$ and $f_U=a_2+b_2+c_2$, and then obtain $[a_1, b_1, c_1]=[-1.2, -0.4, 0.8]$ and $[a_2, b_2, c_2]=[1, -0.5, 0.7]$ by solving (5). Then, based on (3) and (4), we can rewrite (2) as (6), for $\forall j \in \mathcal{M}, t \in \mathcal{T} \setminus \{D\}, i \in \mathcal{N}$.

$$x_{j,i,t+1} \geq x_{j,i,t} + 1.2(v_{j,i,t} - v_{j,i,t+1}) + 0.4(\sum_{i \in \mathcal{N}} v_{j,i,t} - \sum_{i \in \mathcal{N}} v_{j,i,t+1}) - 0.8 \quad (6a)$$

$$x_{j,i,t+1} \leq x_{j,i,t} + (v_{j,i,t} - v_{j,i,t+1}) - 0.5(\sum_{i \in \mathcal{N}} v_{j,i,t} - \sum_{i \in \mathcal{N}} v_{j,i,t+1}) + 0.7 \quad (6b)$$

### B. Constraints of Travel Time

Let us define the matrix $\mathbf{T}_j$, where the element $T_{j,i_1 i_2}$ in the $i_1$th row and $i_2$th column denotes the time spans spent traveling from node $i_1$ to node $i_2$ for MER $j$. $\mathbf{T}_j$ can be predetermined before scheduling the MERs by the algorithms for the shortest path issue, e.g., Dijkstra's algorithm and the Floyd-Warshall algorithm [4], or even the mature map applications. We follow [5]-[9] and assume $\mathbf{T}_j$ is already known. In addition, besides the travel time $T_{j,i_1 i_2}$, if we hope to further consider the time spent by MER $j$ in some other procedures around a travel including, e.g., being disconnected and departing from node $i_1$ (denoted as $T_{j,i_1+}$), being parked at and connected to node $i_2$ (denoted as $T_{j,i_2-}$), we can use the sum $T_{j,i_1+}+T_{j,i_1 i_2}+T_{j,i_2-}$ as the new element in the $i_1$th row and $i_2$th column to compose the matrix $\mathbf{T}_j$.

Then, our next step is to extract $T_{j,i_1 i_2}$ from $\mathbf{T}_j$ when the MER starts traveling from node $i_1$ to $i_2$. Two auxiliary matrixes $\mathbf{A}_{j,t}$ and $\mathbf{B}_{j,t}$ are defined as

$$\mathbf{A}_{j,t} = \left[ x_{j,1,t} \cdot T_{j,1*}^T, x_{j,2,t} \cdot T_{j,2*}^T, \cdots, x_{j,N,t} \cdot T_{j,N*}^T \right]^T \quad (7a)$$

$$\mathbf{B}_{j,t} = \left[ v_{j,1,t} \cdot T_{j,*1}, v_{j,2,t} \cdot T_{j,*2}, \cdots, v_{j,N,t} \cdot T_{j,*N} \right] \quad (7b)$$

where $T_{j,i*}$ and $T_{j,*i}$ represent the $i$th row and the $i$th column of $\mathbf{T}_j$, respectively.

Let us take a look at the matrix $\mathbf{C}_{j,t}=\mathbf{A}_{j,t-1}+\mathbf{B}_{j,t}-\mathbf{T}_j$. When $t=t_1+2$ (i.e., MER $j$ starts traveling from node $i_1$ to $i_2$), we can derive that all the row sums of $\mathbf{C}_{j,t_1+2}$ are below 0 except for that of the $i_1$th row, and the sum of the $i_1$th row is exactly $T_{j,i_1 i_2}$. In other words, the travel time spent is equal to the maximum row sum of matrix $\mathbf{C}_{j,t_1+2}$. For any other $t$ in Table I, we can obtain that the maximum row sum of $\mathbf{C}_{j,t}$ is below or equal to 0. Thus, let us define the variable $S_{j,t}$ for $\forall j \in \mathcal{M}, t \in \mathcal{T} \setminus \{0\}$ as

$$S_{j,t} = \max \left\{ 0 \cup \left\{ x_{j,i,t-1} \cdot \sum_{k \in \mathcal{N}} T_{j,ik} + \sum_{k \in \mathcal{N}}(v_{j,k,t} \cdot T_{j,ik}) - \sum_{k \in \mathcal{N}} T_{j,ik} \middle| i \in \mathcal{N} \right\} \right\} \quad (8)$$

The set after '0' represents all the row sums of $\mathbf{C}_{j,t}$. For the sake of modeling, the relaxed form of (8) can be written as

$$S_{j,t} \geq x_{j,i,t-1} \cdot \sum_{k \in \mathcal{N}} T_{j,ik} + \sum_{k \in \mathcal{N}}(v_{j,k,t} \cdot T_{j,ik}) - \sum_{k \in \mathcal{N}} T_{j,ik},$$
$$\forall i \in \mathcal{N}, j \in \mathcal{M}, t \in \mathcal{T} \setminus \{0\} \quad (9a)$$

$$S_{j,t} \geq 0, \forall j \in \mathcal{M}, t \in \mathcal{T} \setminus \{0\} \quad (9b)$$

In our model, $S_{j,t}$ represents the travel time to be consumed by MER $j$, as shown in Table I. For a specific routing problem of MERs, whether it is desirable to reach a high resilience of the power system or for other purposes, time wasted in travel (i.e., $S_{j,t}$ over $T_{j,i_1 i_2}$) is not expected and is not optimal. This means $S_{j,t}$ will be minimized as much as possible while (9) holds in the solving process, and thus, (9) is equivalent to (8). Then, $S_{j,t}$ is equal to $T_{j,i_1 i_2}$ only when MER $j$ starts traveling from $i_1$ to $i_2$, i.e., when $x_{j,i_1,t-1}=1$ and $v_{j,i_2,t}=1$, for $i_1, i_2 \in \mathcal{N}$. Actually, we can imagine $S_{j,t}$ as the fuel supplemented at time $t$, and the MER is 'refueled' only at the start of each travel.

Then, we define the variable $R_{j,t}$ as

$$R_{j,t} = R_{j,t-1} + S_{j,t} - \sum_{i \in \mathcal{N}} v_{j,i,t-1}, \forall j \in \mathcal{M}, t \in \mathcal{T} \setminus \{0\} \quad (10)$$

Similarly, $R_{j,t}$ can be imagined as the residual fuel of MER $j$ at time $t$. As shown in Table I, $R_{j,t}$ decreases with travel. Additionally, constraint (11) is used to maintain the traveling state until $R_{j,t}$ is 'used up', where $M$ is a large positive number.

$$\frac{R_{j,t}}{M} \leq \sum_{i \in \mathcal{N}} v_{j,i,t} \leq R_{j,t}, \forall j \in \mathcal{M}, t \in \mathcal{T} \quad (11)$$

The binary variable $w_{j,t}$ is defined for $\forall j \in \mathcal{M}$ and $t \in \mathcal{T} \setminus \{0\}$ to maintain the direction of the MER during each travel as follows.

$$w_{j,t} \geq \sum_{i \in \mathcal{N}} v_{j,i,t-1} + \sum_{i \in \mathcal{N}} v_{j,i,t} - 2 + \varepsilon, \forall j \in \mathcal{M}, t \in \mathcal{T} \setminus \{0\} \quad (12a)$$

$$-(1-w_{j,t}) \leq v_{j,i,t} - v_{j,i,t-1} \leq (1-w_{j,t}), \forall i \in \mathcal{N}, j \in \mathcal{M}, t \in \mathcal{T} \setminus \{0\} \quad (12b)$$

where $\varepsilon$ is a small positive number and $0 < \varepsilon \leq 1$.

The constraints for the initial conditions are written as (13), where $i_j$ represents the initial node where MER $j$ is parked.

$$x_{j,i_j,0} = 1, S_{j,0} = 0, R_{j,0} = 0, w_{j,0} = 0, \forall j \in \mathcal{M} \quad (13)$$

The elaboration about the derivation in the above two sections has been uploaded as [10] by the authors. The final form of the proposed mobility model contains (1), (6), (9)-(13), and the variables involved includes the binary ones $x_{j,i,t}, v_{j,i,t}, w_{j,t}$, and the continuous ones $S_{j,t}, R_{j,t}$.

### C. Estimation of the Size of the Models

We estimate the size of the model proposed in this paper and three other representative models, as shown in Table III, where $N$ is the number of nodes that support the connection of MER, i.e. $N=|\mathcal{N}|$, $N_v$ is the number of virtual nodes introduced in TSN [5], [6], and $M$ is the number of MERs, i.e., $M=|\mathcal{M}|$. In this estimation, we assume the same matrix $\mathbf{T}_j$ for any $j \in \mathcal{M}$ and the $i$th-row and $k$th-column element is written as $T_{ik}$. The elaboration of the estimation can be found in [10]. Specially, for the SWBM, if we further assume all $T_{i_1 i_2}$ are equal to a specific value $T_{av}$, we can estimate the minimum number of constraints when $T_{av}=1$, which is $M[D(N^2-N)+2D+2]$. This number increases as $T_{av}$ increases. The results in Table III show that the proposed model eliminates the drawback in existing mobility models that the number of variables or constraints increases with the square of nodes. This can lead to better efficiency especially when we route MERs among many nodes using a small time span, and we will demonstrate that later.

## III. NUMERICAL RESULTS

### A. A Simple Programming for Testing

For testing, we construct a simple MILP model by (14) to schedule MERs for electric service restoration after the power grid suffers great faults. Communication between MERs and decision maker (e.g., the power system operator) can be realized in the wireless mode through public, private, or satellite networks. Here we aim to restore electric energy as much as possible for the interrupted customers while considering the

TABLE III
COMPARISON OF THE MODEL SIZE AMONG THE MOBILITY MODELS

| Mobility Models | Number of binary variables | Number of constraints |
|---|---|---|
| Proposed model, (1), (6), (9)-(13) | $M(D+1)(2N+1)$ ( and $2M(D+1)$ continuous variables in addition. ) | $MD(5N+6)+7M$ |
| TSN, (1)-(4) in [5] | $DM(N^2+2N_v)$, where $N_v=\sum_{i=1}^{N-1}\sum_{k>i} T_{ik}-N(N-1)/2$ | $DM(N^2+3N_v+1)-M(N^2-N+2N_v)$ |
| Modified TSN, (4)-(6) in [9] | $M[N^2(D+1)-\sum_{i=1}^{N}\sum_{k=1}^{N} T_{ik}-N]$ | $MD(N+1)$ |
| SWBM, (23), (25), (26) in [8] | $M(D+1)(N+1)$ | $M[(2D+1)\sum_{i=1}^{N}\sum_{k=1}^{N} T_{ik}-\sum_{i=1}^{N}\sum_{k=1}^{N} T_{ik}^2+4D+4]/2$ |

costs of MERs by traveling, as formulated by

$$\max \sum_{t \in \mathcal{T}} \left[ \sum_{l \in \mathcal{L}} \left( y_l(t) \cdot \sum_{i \in \mathcal{N}_l} W_i \cdot P_i(t) \right) \right] \cdot \Delta t - \sum_{j \in \mathcal{M}} C_{j, \Delta t} \cdot D_j \quad (14)$$

s.t.: Equations $(1),(6),(9)-(13),(15)$

The first term in (14) is the energy restored by MERs. Suppose that the power grid is separated by faults into several islands, represented by set $\mathcal{L}$, and recovery of the faults is considered. In addition to the mobility model, a full programming model of scheduling MERs for some purposes, e.g., service restoration [5], [8], [9], economic dispatch [6], also includes the constraints to restrict the operation of power grid and MERs, however, they are not the concern of this paper and are compressed as much as possible in this test. Thus, for simplicity and to focus on the performance of the mobility model itself, the operational constraints of the power grid and MERs themselves are ignored in the test, which means 1) all of the nodes in one island can be restored as long as an MER is parked at any node in it and 2) the MERs are not limited in power or energy. $\mathcal{N}_l$ represents the set of nodes in island $l$. $P_i(t)$ and $W_i$ are the interrupted load and its weight for node $i$. The auxiliary binary variable $y_l$ is used to indicate the restored state of island $l$, and some relevant constraints should be added into (14) based on assumption 1) for the specific mobility model. For example, for our proposed model, they can be written as

$$\frac{\sum_{j \in \mathcal{M}} \sum_{i \in \mathcal{N}_l} x_{j,i,t}}{M} \leq y_l(t) \leq \sum_{j \in \mathcal{M}} \sum_{i \in \mathcal{N}_l} x_{j,i,t}, \forall l \in \mathcal{L}, t \in \mathcal{T} \quad (15)$$

The second term in (14) represents the travel costs of MERs, and this term can prevent useless travel of MERs. $C_{j,\Delta t}$ is the energy consumption of MER $j$ by traveling for one time span, and we can always convert it into 'kW·h' based on the fuel and electric prices so that the two terms in (14) have the same units. $D_j$ is the total travel spans of MER $j$. For our model, $D_j$ can be represented by $\sum_{t \in \mathcal{T}} \sum_{i \in \mathcal{N}} v_{j,i,t}$.

### B. Test and Comparison Results

We perform the test based on the modified IEEE 37-node test feeder [11]. In addition to the model proposed in this paper, the three other representative models, the general TSN in [5], the modified TSN in [9], and the SWBM in [8], are chosen for comparison in this test, and we just need to replace '(1), (6), (9)-(13)' with the other models and modify (15) to conduct the programming of (14). The test is implemented using MATLAB R2018b with YALMIP toolbox, and the programming is solved by Gurobi 9.0.0 on a computer with an Intel Core i5-8250U processor and 12 GB of memory. The MIP gap is set to $1\times10^{-5}$.

The scheduling horizon is taken as 6 h, and two MERs are adopted here. The two MERs are supposed to be the same and be driven by electric power, with a consumption of 0.3 kW·h for 10 min. Their speed is set to 1000 ft./min. Four faults (at lines 5-9, 3-6, 23-26, and 33-34) are supposed to occur at the same time, *i.e.*, at the initial moment of the 1st span, and be repaired 70 min, 130 min, 230 min and 320 min later, respectively. The scheduling result using our proposed model for the total 37 nodes with a 10-min time span is illustrated in Fig. 1. The results of the four models have the same optimal value of the objective function, *i.e.*, $7.58\times10^3$ kW·h, though the optimal routes are slightly different.

For a comparison of the model size and computational efficiency of the four models, we also perform a series of tests under different values of the time span and the number of nodes that support the connection of MERs. The results are summarized in Fig. 2. The proposed model always has a smaller size than the other models from a comprehensive perspective. The optimal value of the objective function decreases with an increase in the chosen time span, which is $7.58\times10^3$ kW·h for a 10-min span, $6.70\times10^3$ kW·h for a 20-min span, and $6.22\times10^3$ kW·h for a 30-min span. This shows that a smaller time span seems more beneficial to the quality of the routing of MERs. Moreover, for most situations, the proposed model consumes the shortest time to obtain the optimal result, and this merit in terms of computational efficiency becomes increasingly apparent as the number of nodes increases and the length of the time span decreases. Especially under the situation where all the 37 nodes and a 10-min time span are chosen, as shown in Fig. 2(l), the time consumption of the proposed model is only approximately one tenth of that of the general TSN.

## IV. Conclusion

In this paper, we focused on the routing issue of MERs, and proposed a novel mobility model to support it. The model was derived around two key points of the behavior of MERs: the state transitions and the necessary travel time between nodes, and all the constraints composing the model were formulated in linear form, which can be well embedded in programming for routing of MERs. The quantitative estimation showed the advantage of the proposed model in size. The results of test further validated the better computational efficiency of the proposed model than the others, especially under the cases where MERs were routed among many nodes with a small time step. The proposed model can be well recommended to solve the routing and scheduling issues of MERs.

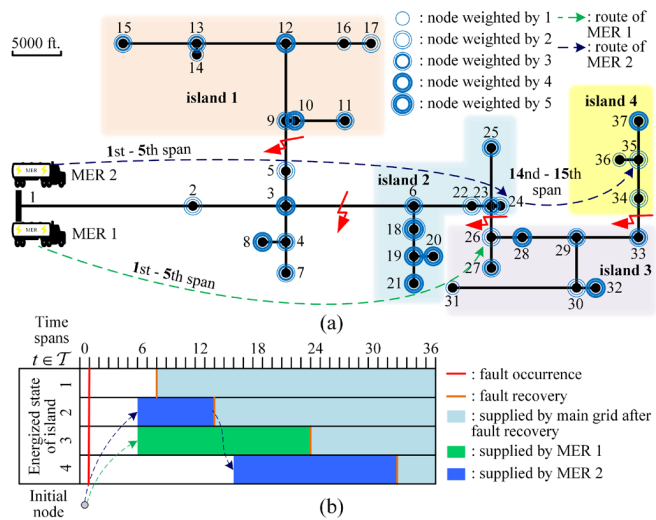

Fig. 1. (a) illustrates the routing of two MERs among 37 nodes with a 10-min time span. (b) is the energized state of the four islands.

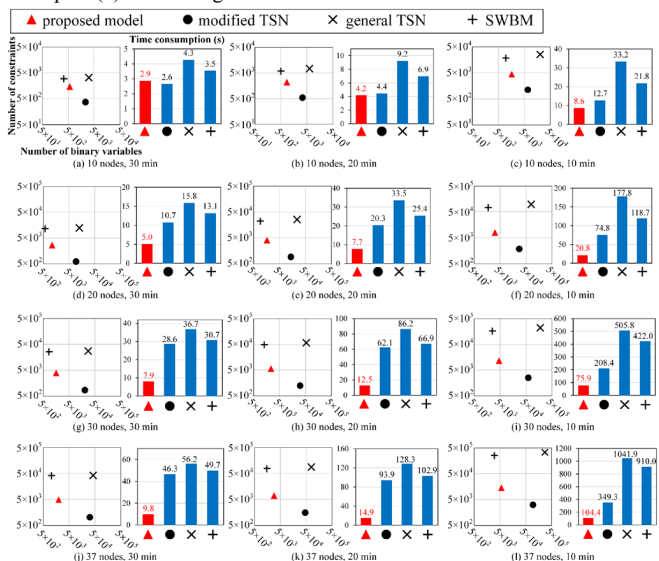

Fig. 2. Comparison of the model size and computational efficiency among the proposed model and the other three. The logarithmic coordinate is adopted for the numbers of variables and constraints.